\documentclass{vgtc}                          

\usepackage{mathptmx}
\usepackage{graphicx}
\usepackage{times}
\usepackage{enumitem}
\usepackage{url}
\let\ifpdf\relax
\usepackage[hidelinks]{hyperref}

\onlineid{0}

\vgtccategory{Research}

\vgtcinsertpkg

\title{A Web-based Large-scale Timelapse Editor for Creating and Sharing Guided Video Tours and Interactive Slideshows}

\author{Yen-Chia Hsu\thanks{e-mail: yenchiah@andrew.cmu.edu}\qquad %
		Paul Dille\thanks{e-mail: pdille@andrew.cmu.edu}\qquad %
		Randy Sargent\thanks{e-mail: rsargent@andrew.cmu.edu}\qquad %
		Christopher Bartley\thanks{e-mail: cbartley@andrew.cmu.edu}\qquad %
		Illah Nourbakhsh\thanks{e-mail: illah@andrew.cmu.edu}\\ %
        \scriptsize The Robotics Institute, Carnegie Mellon University }

\abstract{
Scientists, journalists, and photographers have used advanced camera technology to capture extremely high-resolution timelapse and developed information visualization tools for data exploration and analysis. However, it takes a great deal of effort for professionals to form and tell stories after exploring data, since these tools usually provide little aids in creating visual elements. We present a web-based timelapse editor to support the creation of guided video tours and interactive slideshows from a collection of large-scale spatial and temporal images. Professionals can embed these two visual elements into web pages in conjunction with various forms of digital media to tell multimodal and interactive stories.
} 

\CCScatlist{ 
  Information visualization, Visualization systems and tools
}

\begin{document}


\firstsection{Introduction and Related Work}

\maketitle

As camera technology proliferates, the quantity and resolution of digital images have increased exponentially. Researchers have worked on creating tools for generating, exploring, and sharing large-scale timelapses after capturing high quality images. For instance, Sargent et al.~\cite{Sargent2010} have developed an integrated solution to capture and stitch gigapixel timelapses, generate multi-resolution video tiles, and visualize the results in an interactive web-based viewer. Professionals have used the technology to document the entire context of a site, capture extreme details in scenes, and share high resolution timelapses on the Internet. One example is the Google Annual Earth Timelaspe~\cite{EarthEngine} consisting of 29 cloud-free mosaics of the planet from Landsat satellite imagery between 1984 and 2012, with each frame containing nearly 1 trillion explorable pixels.

Such visualization tools enable browsing large-scale images and provide powerful data exploration experiences to professionals. However, most existing tools lack the capability for professionals to create visual elements for data-driven storytelling through space and time, affording the presentation of a sequence of related facts found during exploration. If professionals want to create video tours, they need to import timelapses into video editing software. This process is impractical and takes a huge amount of time as most software cannot handle large datasets that do not fit into memory.

To address this problem, we present a web-based timelapse editor~\cite{Editor} operating along large-scale space and time to assist professionals in creating visual elements based on facts found during exploration. The editor allows the creation of \textit{guided video tours} and \textit{interactive slideshows} to enhance different story structures described by Segel and Heer~\cite{Segel2010}. Guided video tours present changes over time in author-driven stories, having linear visualization paths and limited interactivity. Interactive slideshows store various interesting locations and facilitate follow-up exploration for reader-driven stories, having little prescribed orderings and high interactivity. Each tour or slideshow is a micro-story and can be integrated by professionals with other types of media into a mega-story~\cite{Jain2013}.

\section{Creating Video Tours or Interactive Slideshows}

\begin{figure}
	\centering
	\includegraphics[width=1\columnwidth]{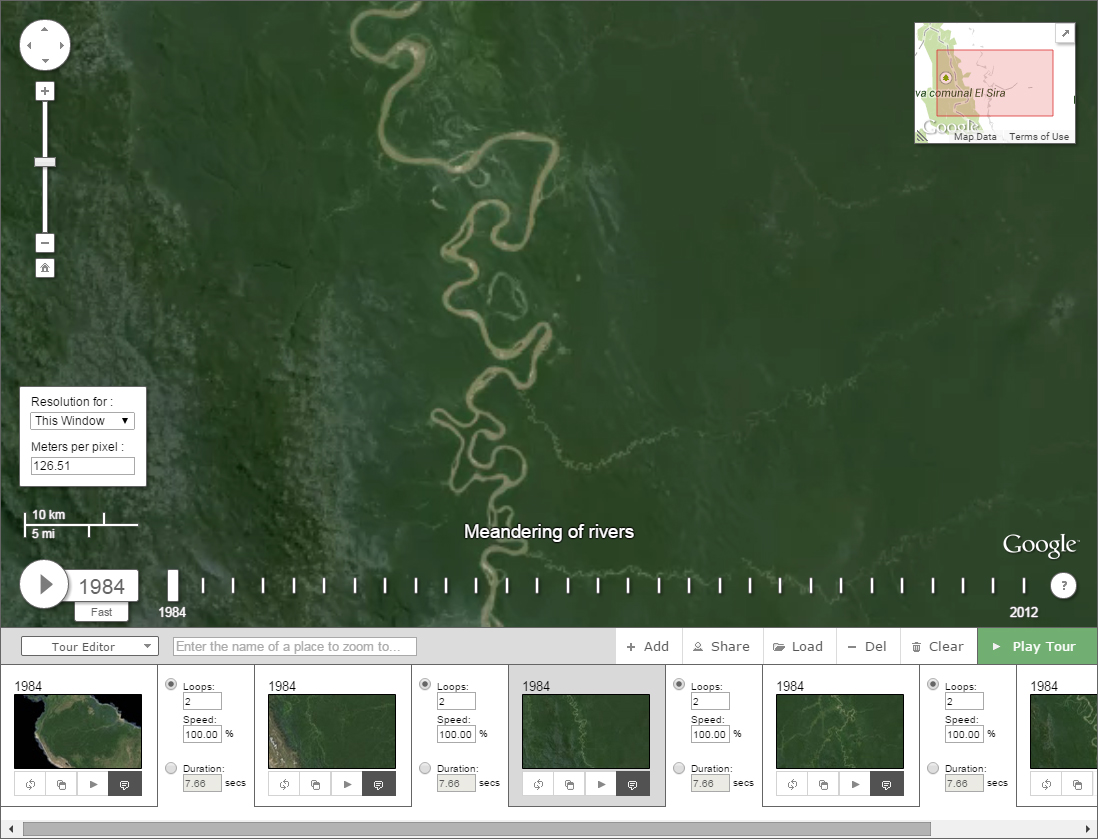}
	\caption{The timelapse editor.}
	\label{fig:editor}
\end{figure}

When scientists find interesting events while exploring the timelapse in the zoomable and pannable viewer, they can use the editor attached at the bottom of the viewer (Figure~\ref{fig:editor}) to create guided video tours defined by sequences of keyframes containing the time, location, and scale of different views. Users can use the functions provided on the main control bar to add keyframes. On the left side of the viewer above the scale bar, there is a small box displaying the satellite image quality relative to a resolution to assist scientists in choosing appropriate scales. If a keyframe is unwanted or misplaced, users can select the keyframe and then delete it or drag it to a desired place in the sequence. Each keyframe in the container has auxiliary functions for users to update the keyframe to the current view, duplicate the keyframe, and add corresponding descriptions.

Users can click on the play button on the main control bar to preview the tour animated by using a linear motion for each pair of keyframes using default transition settings. In the keyframe container, users can specify transition parameters between two consecutive keyframes. There are two main types of transitions: speed and duration. A speed transition uses the user-defined playback rate relative to the original video rate and automatically computes the duration. The editor uses 100\% speed as the default transition setting. In contrast, a duration transition calculates the speed accordingly from the user-defined duration. For short timelapses (e.g. less than 100 frames), users can assign a desired looping parameter, the number of times to loop through the entire timelapse video between two keyframes. While looping the entire video, the editor introduces a 0.5 second dwelling time for a better transition effect, meaning that the animation pauses at the very beginning and end of the timelapse for 0.5 second. By using the animation logic described above, users can perform the following five different camera motions:

\vspace{-0.2em}
\begin{itemize}[leftmargin=*]
	\itemsep-0.3em
	\item
	Animate zooming, panning, and time simultaneously by adding two keyframes at different locations and dates, and then setting speed or duration to a non-zero value.
	\item
	Pause zooming and panning but animate time by adding two keyframes at different dates but the same location, and then setting speed or duration to a non-zero value.
	\item
	Pause time but animate zooming and panning by adding two keyframes at different locations but the same date, and then setting duration to a non-zero value.
	\item
	Pause zooming, panning, and time simultaneously by adding two keyframes cloned at the same location and date, and then setting duration to a non-zero value.
	\item
	Jump immediately from the first keyframe to the second one by forcing duration to be zero.
\end{itemize}

\begin{figure}
	\centering
	\includegraphics[width=1\columnwidth]{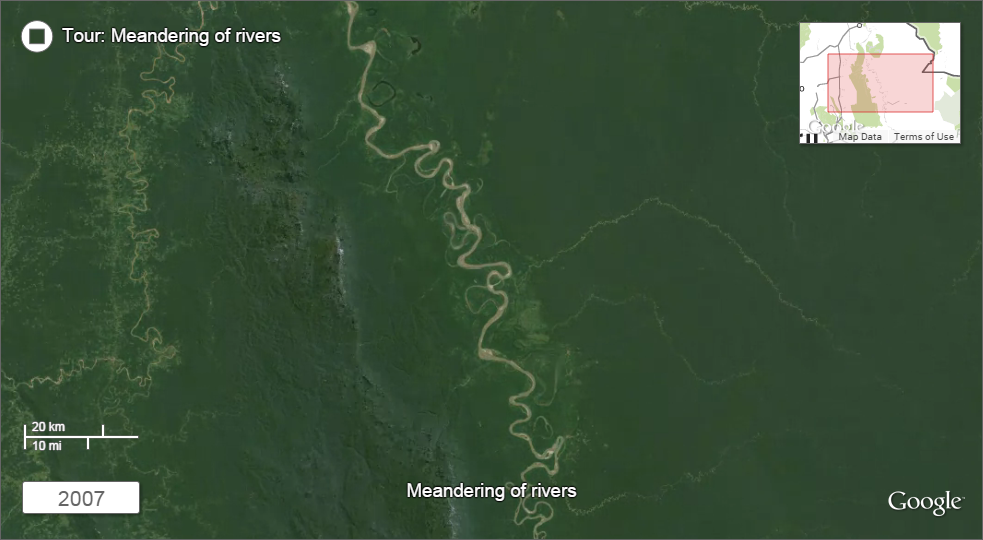}
	\caption{The guided video tour.}
	\label{fig:tour}
\end{figure}

When finished editing, users can click the share button to disseminate or embed the guided video tour (Figure~\ref{fig:tour}) encoded in an URL (uniform resource locator) into a storytelling webpage. Descriptions associated with each keyframe show up as video captions. The tour interface displays a time stamp, a scale bar, and a small Google map to provide contextual information. A button at the top left of the interface allows users to stop the tour.

\begin{figure}
	\centering
	\includegraphics[width=1\columnwidth]{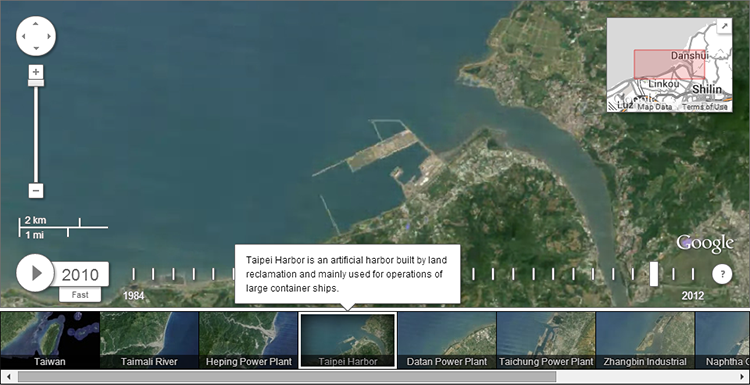}
	\caption{The interactive slideshow.}
	\label{fig:slideshow}
\end{figure}

Professionals can also use the editor to create interactive slideshows (Figure~\ref{fig:slideshow}) containing a collection of  locations. The workflow is similar to the one for creating video tours. The editor turns keyframes into slides and omitting all transition parameters. Audiences first see an overview and then can choose an interesting location to zoom in and to request detailed information. When audiences hover a mouse onto a slide, a message box containing corresponding descriptions fades into the interface. Clicking on a slide animates the viewer to a keyframe representing an interesting location. Professionals can use interactive slideshows for storytelling in a webpage or for visual exhibitions on hyperwalls in museums.

\section{Discussion and Future Work}
After releasing the editor in 2012, we collaborated with a journalist at TIME magazine in telling a story about extreme natural resources, global climate changes, and urban explosions on Earth. We render the tours into high-quality videos and then the journalist integrated these videos with the timelapse viewer and other forms of digital media into a compelling story~\cite{EarthTime}. Audiences first experienced a prescribed author-driven story with rich multimedia containing space-time tours and then were free to explore the timelapse by themselves. We also worked with scientists in the Explorables team in Pittsburgh in 2014 to create interactive slideshows for telling a reader-driven story about landscape changes along Taiwan's coastline over a two-decade period~\cite{Explorables}. In 2014 we assisted scientists at the Exploratorium museum in San Francisco to install a visual exhibition showing an interactive slideshow on a hyperwall. The professionals we collaborated with were able to create tours or slideshows by using the editor over approximately half an hour of training and to use these visual elements in forming interesting stories. However, there are research questions we need to answer:

\vspace{-0.2em}
\begin{itemize}[leftmargin=*]
	\itemsep-0.3em
	\item
	What is the efficiency of the editor? Can professionals use it easily without spending too much effort?
	\item
	Do stories integrating custom guided tours and interactive slideshows encourage audiences to explore the timelapse?
	\item
	Do custom guided tours and interactive slideshows help audiences gain insights from stories formed by professionals?
\end{itemize}

Developing tools to support the creation of visual elements for storytelling depends heavily on the user needs from professionals. It is vital to collaborate with target users and keep them in the design loop. In the future, we plan to conduct a medium to long term user evaluation~\cite{Shneiderman2006} to answer the research questions. We hope the editor can truly help professionals focus more on the content of stories rather than time-consuming and laborious work.

\acknowledgements{
	Google Earth Engine, TIME magazine, and all other participants.
}

\bibliographystyle{abbrv}
\bibliography{reference}
\end{document}